Bulky counterions: enhancing the two-photon excited fluorescence of gold nanoclusters.

Franck Bertorelle[a]‡, Christophe Moulin[a]‡, Antonin Soleilhac[a], Clothilde Comby-Zerbino[a], Philippe Dugourd[a], Isabelle Russier-Antoine[a], Pierre-François Brevet[a] and Rodolphe Antoine[a]*

Abstract: Increasing fluorescence quantum yields of ligand-protected gold nanoclusters has attracted wide research interest. The strategy consisting in using bulky counterions has been found to dramatically enhance the fluorescence. In this communication, we push forward this concept to the nonlinear optical regime. We show that by an appropriate choice of bulky counterions and of solvent, a 30-fold increase in two-photon excited fluorescence (TPEF) signal at ~600 nm for gold nanoclusters can be obtained. This would correspond to a TPEF cross section in the range of 0.1 to 1 GM.

The quest for high resolutions and penetration depths with spectroscopic tools has led to innovative approaches in optical microscopy of biological systems. Two-photon excited fluorescence (TPEF) microscopy is probably the most important advance in optical microscopy of biological systems since the introduction of confocal imaging.1 The successful application of TPEF is partly due to the fact that commonly used fluorescent molecules can be efficiently excited by the simultaneous absorption of 2 photons.2, 3 However, to reach ultrahigh detection sensitivity, newly engineered nonlinear optical (NLO) chromophores, endowed with large two-photon absorption cross-sections, have to be developed.

Liganded silver and gold quantum clusters represent a promising direction in the quest to develop novel nonlinear optical nanomaterials. Pioneering studies have clearly demonstrated that ligated noble metal nanoclusters (NCs), with few to hundred atoms exhibit remarkable optical nonlinearities.4-6  Shortly afterwards, strong two-photon emission was reported from uncharacterized gold clusters, making them a promising alternative as contrast agents for live cells imaging.7 Thus, these ligated nanoclusters may provide a valuable route within the nonlinear optical regime. We have described in our combined experimental and theoretical investigations how to improve two-photon absorption (TPA) properties in such quantum nanoclusters, through the concept of "ligand-core" NLO-phores.8 The structure of the metallic core, its charge and symmetry (influencing the NLO cross sections), and the ligands stabilizing the core play a major role in TPA efficiencies.9 However, the main conclusion so far is that liganded silver and gold quantum clusters are excellent two-photon absorbers but are rather poor two-photon excited emitters.10 For example, the two-photon absorption cross-section at 780 nm for Au15 is ~65 700 GM, while its TPEF cross section at 475 nm is only ~0.022 GM.11

Concerning nonlinear emissive properties, both experimental and theoretical progresses are needed in order to produce not only large TPA but also large TPEF cross sections. On the one hand, alloying is a promising strategy. Indeed, Le Guével et al12 reported enhanced fluorescence of gold nanoclusters by silver doping. Likewise, two-photon excited fluorescence emission was also found to be enhanced.13 Unfortunately, the exact stoichiometry of such AuAg nanoclusters remains unknown and has prevented the report of any quantitative estimation of TPEF cross-sections. On the other hand, it is well known that the ligand shell rigidity14, 15 influences the emissive properties of Au NCs. A simple way to increase rigidity of the protective shell is to use as ligands proteins instead of thiolated molecules. Xie et al.16 have developed a facile, one-pot, "green" synthetic technique based on bovine serum albumin (BSA). It was shown that BSA-Au25 clusters also exhibit an efficient two photon absorption followed by red to near-infrared luminescence.17 However, no two photon absorption cross section was reported due to the difficulty in calculating the precise concentration of the clusters in the solution. Using (bio)organic thiolate templates, further enhancement can be achieved by the increased rigidity of the metal-thiolate shell. Pyo et al.18 have shown that it is possible to achieve one-photon excited luminescence quantum yields >60% by rigidifying the gold shell with the binding of bulky groups (e.g. tetraoctylammonium (TOA) cations). In this communication, we push forward this concept to the nonlinear optical regime. We show that by an appropriate choice of bulky counterions and solvents, a dramatic enhancement of TPEF cross sections for Au15(SG)13 and Au18(SG)14 can be obtained.

Glutathione, the ligand molecule used in this work to protect the AuNCs, possesses carboxylate groups that can efficiently bind with bulky ammonium cations.18 Two quaternary ammonium cations with different chain lengths were explored in this work, namely tetrabutylammonium (TBA) and tetraoctylammonium (TOA), see Fig. S1 in supporting information. We performed negative-mode electrospray ionization mass spectrometry (see Figure S2 in supporting information showing the negative-mode ESI mass spectrum of TBA-Au15 and TBA-Au18 clusters). There are peaks observed at selected m/z windows that represent TBA-Au15 and TBA-Au18 ions with the overall charge of 5– and 7- respectively. Additional peaks after the bare Au NC ions correspond to different number of counterions (TBA+) paired with the carboxylate of the glutathione ligands. This indicates that the number of TBA cations paired with Au NCs gas phase ions up to 7. Also, as can be seen in Figure S2, the TBA-Au15 and TBA-Au18 cluster ions retain their original respective Au15(SG)13 and Au18(SG)14 stoichiometry. This is in agreement with the previous report of Pyo et al.18 who found indication that the number of TOA cations paired with Au22(SG)18 was >13. Absorption spectra of Au NCs and paired bulky cations-Au NCs are almost identical. In stark contrast, as already observed by Pyo et al.18, the one-photon excited fluorescence (OPEF) intensity was drastically enhanced after pairing Au NCs with bulky ammonium cations in methanol (see Fig. 1a). The intensity increased 10-fold for TBA-Au15 clusters in methanol (as compared to Au15 clusters in water). The use of methanol and of bulky counterions further inhibit loss channels specific to water. Time-resolved luminescence measurements were carried out for paired bulky cations-Au15 NCs and showed a drastically increased lifetime for TBA-Au15 compared to that for Au15 clusters (see Figure S3 in supporting information). The 2-photon excited fluorescence (TPEF) spectra are reported for Au NCs and for paired bulky cations-Au NCs (see Fig. 1b). As already observed for the linear regime (OPEF), their TPEF intensity was drastically enhanced after pairing Au NCs with bulky ammonium cations in methanol. The TPEF intensity increased ~10-fold for TBA-Au15 clusters in methanol. TBA is the best candidate to enhance (both one-photon and two-photon excited) luminescence properties of Au NCs. With TPEF (see Fig. 1b), the red emission is much weaker above 700 nm than the one observed with OPEF (compare with Fig. 1a). This is due to the fact that we used for the 2PL setup a low pass filter (~680 nm), in order to remove the photons coming from the excitation laser beam. Of note is that TPEF spectra of bulky cations-Au15 NCs, exhibiting emission bands in the blue and the red, contrast with linear photoluminescence (OPEF) spectra where emission only in the red is observed (compare Fig. 1a and 1b). Such a difference between emission resulting from absorption of one or two photons was already reported on DNA-protected6 and thiolate-protected silver clusters.19 This behavior might be accounted for the different nature of the metal-to-metal excitations within the gold cluster core (blue emission) and the ligand-metal and ligand-metal-metal charge transfer excitations (red emission).8, 9, 19

We also explored the solvent dependence of TPEF. The TPEF intensity of both paired bulky cations, Au15 and Au18 NCs, significantly increases with decreasing dielectric constant of the solvent (methanol versus DCM, see Fig. 2). Indeed, more polar is the medium, weaker is the effect of ion pairing. In DCM, with pairing with TBA cations, the TPEF intensity around 600 nm increases ~30-fold as compared to Au15(SG)13 in water. This would correspond to a TPEF cross section in the range of 0.1 to 1 GM. A similar enhancement is observed for Au18(SG)14. However, for Au18(SG)14 TPEF spectra, the relative ratio between emission bands in the blue and the red seems to depend on the solvent.

In order to evaluate the use of both paired bulky cations-Au NCs as possible contrast agents for multiphoton imaging, we investigated methanol droplets (formed in heptane) doped with the nanoclusters using 2PEF microscopy. Figure S4 presents XY (TPEF) intensity scan of selected droplets. The experimental conditions were optimized in order to obtain good signal-to noise ratio and to avoid photobleaching and photodamage of the sample. The fluorescence microscopy observation of NCs doped droplets reveals that the nanoclusters are spread within the entire layer of the sample. A clear image of the droplets by TPEF microscopy is obtained and correlates well with the corresponding optical image. With the advantage of the intrinsic 3D resolution of the multiphoton microscope, we performed several scans in the same region of the droplets, but with different Z positions (see Fig. 3). It was found that Au NCs accumulated within the whole volume of the methanol droplets.

In summary, in this work, we push forward the strategy consisting in using bulky counterions to dramatically enhance the luminescence in the nonlinear optical regime. We show that by an appropriate choice of bulky counterions (tetrabutyl ammonium) and of solvent (dichloromethane), a 30-fold increase in two-photon excited fluorescence signal in the red for Au15(SG)13 can be obtained. This allowed us to record 2-photon images of doped droplets. However, the TPEF cross sections for gold nanoclusters remain lower than common organic fluorophores exhibiting cross sections in the range 10-30 GM.[20, 21] Although glutathione protected gold nanoclusters are known to present exceptional photostability (as compared to usually poor photostability for organic dyes),[7] stronger TPEF emissive cross sections are highly desirable. Other strategies for increasing the TPEF quantum yield need to be addressed. For instance the combination of bulky counterions with other strategies consisting in reinforcing the protecting template (with proteins for example) or using gold core-silver doping has to be demonstrated. Works under these lines are currently undertaken in our lab.

Experimental Section

Materials. Gold(III)chloride trihydrate (HAuCl4,3H2O), was purchased from Alfa Aesar. Reduced L-glutathione (GSH), glacial acetic acid and tributylamine were purchased from Carl Roth. Methanol (meOH), ethanol (Et2O), dichloromethane (DCM), diethylether, tetrabutylammonium borohydride, tetraoctylammonium bromide were purchased from Sigma-Aldrich. Ultrapure water with a resistance of 18.2 MΩ was used throughout the experiment. All of the chemicals were used as received without further purification.

Cluster synthesis. Au15(SG)13 was synthesized as reported by Russier-Antoine et al.[11] Au18(SG)14 was synthesized as follow: 390 mg of glutathione (GSH) is dissolved in 45 ml of methanol, 8 ml of tributylamine and 4.5 ml of water. Then 2 ml of a water gold solution (200 mg of HAuCl4,3H2O) and 20 ml of diethyl ether are added. The solution is stirred in an ice bath for 15 min. 80 mg of tetrabutylammonium borohydride powder is added in two parts (2 x 40 mg) spaced from 45 min, under strong agitation. 45 min later, the yellow solution is removed from the ice bath and 200 mg of tetramethylammonium borohydride is added under strong agitation during 3 hours.

NCs purification. Precipitation of NCs is induced by adding 1 ml of NaOH (1M). The Unwanted products are removed with cycles of dissolution/precipitation. The powder is dissolved in a minimum of slightly basic H2O and precipitated with MeOH. After centrifugation, the powder is dissolved again in 10 ml of water. To remove sodium salt, 2 ml of ice cold acetic acid is added and solution is left undisturbed 1 hour before being centrifuged. The supernatant is collected and then precipitated with MeOH before being dry under vacuum.

TBA-, and TOA-AuNCS preparation. AuNCs (50 mg) are dissolved in 1 ml of water with 200 mg of tetrabutylammonium hydroxide. Then 2 ml of methanol is added and bulky ions-AuNCS are precipitated with diethyl ether. Cycles of dissolution (MeOH)/precipitation (Et2O) were done to remove excess of ammonium salts. TOA-AuNCS were prepared as reported by Pyo et al.18 using Au15(SG)13 and Au18(SG)14 instead of Au22(SG)18.

Optics. One-photon Fluorescence set-up. Fluorescence lifetime was measured on a custom-built set-up as described in refs.22, 23 The fluorescence lifetimes were obtained using a bi-exponential fitting model on the fluorescence decays. The steady-state fluorescence spectra were recorded by an Ultracompact spectrophotometer Econic (B&WTek Inc., Newark, DE, U.S.A.) with a resolution of 1.5 nm via an optical fiber using the same excitation at 500 nm and appropriate optical filters. A low pass Fast Fourier Transform algorithm was then applied to the spectra in order to smooth the signal.

Two-photon Fluorescence set-up. Two-photon fluorescence measurements were performed with a customized confocal microscope (TE2000-U, Nikon Inc.) in which the excitation light entrance has been modified to allow free-space laser beam input, instead of the original optical-fiber light input. The luminescence was excited at 780 nm with a mode-locked frequency-doubled femtosecond Er-doped fiber laser (C-Fiber 780, MenloSystems GmbH). The laser beam was focused by a Nikon CFI Plan Apo VC objective (20x/0.75 NA). The sample was XY scanned by the inner microscope galvanoscanner (confocal C1 head, Nikon Inc.), and the Z scan was performed by the inner microscope motorized focus. The emitted signal was collected in epifluorescence illumination mode. The two-photon fluorescence emission was separated from the incident light through a dichroic mirror and detected by a -20°C cooled photomultiplier tube (R943-02, Hamamatsu Photonics). A FF01-680/SP filter was used in order to remove the photon coming from the excitation laser. The two-photon fluorescence emission spectra were recorded using an iHR320 spectrometer equipped with a 53024 grating from Horiba Jobin Yvon. The initial output power of the femtosecond laser was 62 mW. For TPEF imaging, we reduced this power with OD filters so that only 5 mW were focused, so as not to damage the sample.

COMMUNICATION

Chain my heart. By an appropriate choice of bulky counterions and of solvent, a 30-fold increase in two-photon excited fluorescence signal at ~600 nm for gold nanoclusters can be obtained.

Franck Bertorelle, Christophe Moulin, Antonin Soleilhac, Clothilde Comby-Zerbino, Philippe Dugourd, Isabelle Russier-Antoine, Pierre-François Brevet and Rodolphe Antoine*

*Page No. – Page No.

Bulky counterions: enhancing the two-photon excited fluorescence of gold nanoclusters.

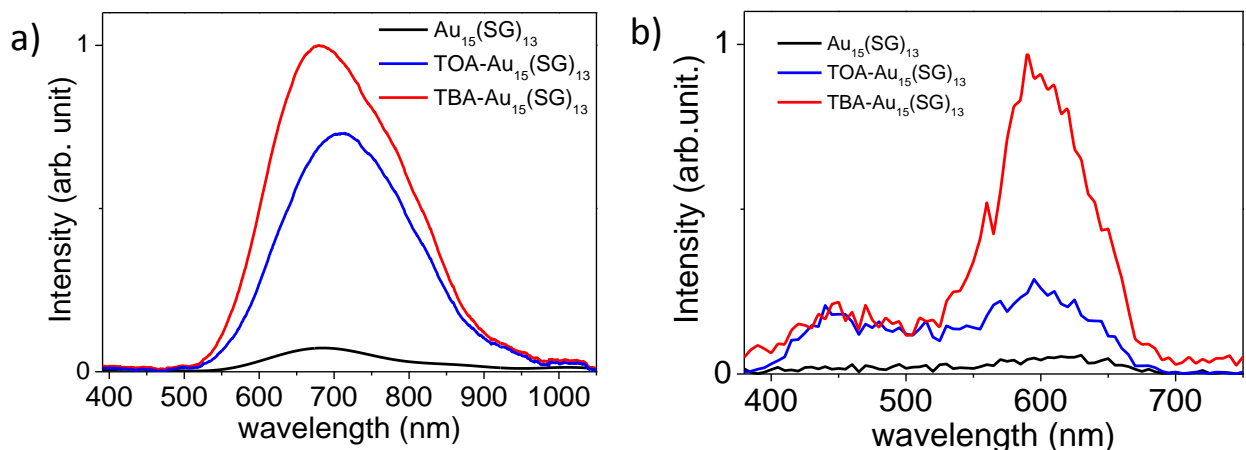

**Figure 1**: a) One photon excited fluorescence spectra of $Au_{15}(SG)_{13}$ in water, TOA-$Au_{15}(SG)_{13}$ and TBA-$Au_{15}(SG)_{13}$ in methanol. Cluster solutions have the same absorbance (0.01) at 500 nm. b) Two photon excited fluorescence spectra at excitation wavelength 780 nm of $Au_{15}(SG)_{13}$ in aqueous

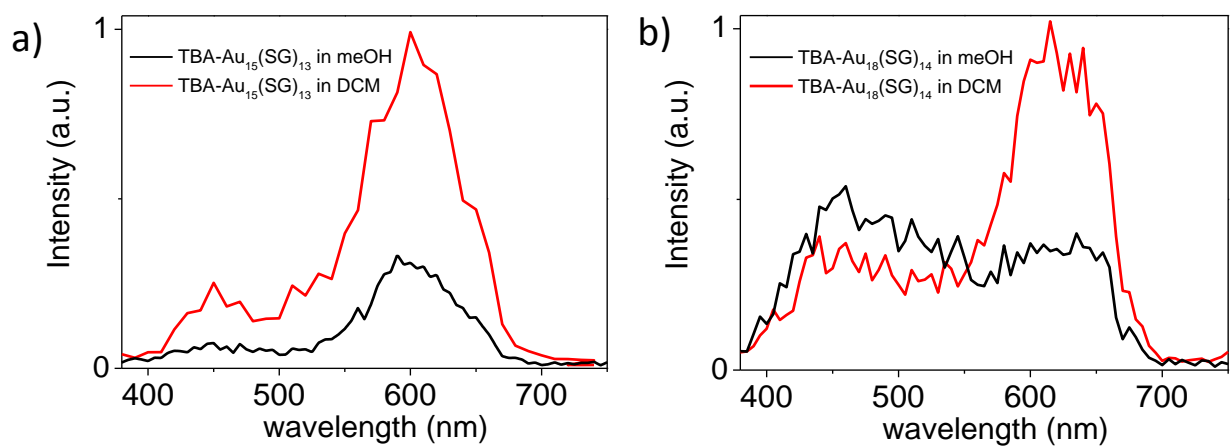

**Figure 2 :** Two photon excited fluorescence spectra at excitation wavelength 780 nm of (left) TBA-$Au_{15}(SG)_{13}$ and (right) TBA-$Au_{18}(SG)_{14}$ in methanol (meOH) and in dichloromethane (DCM) (with the same Au NCs concentration, 2 mg.mL$^{-1}$).

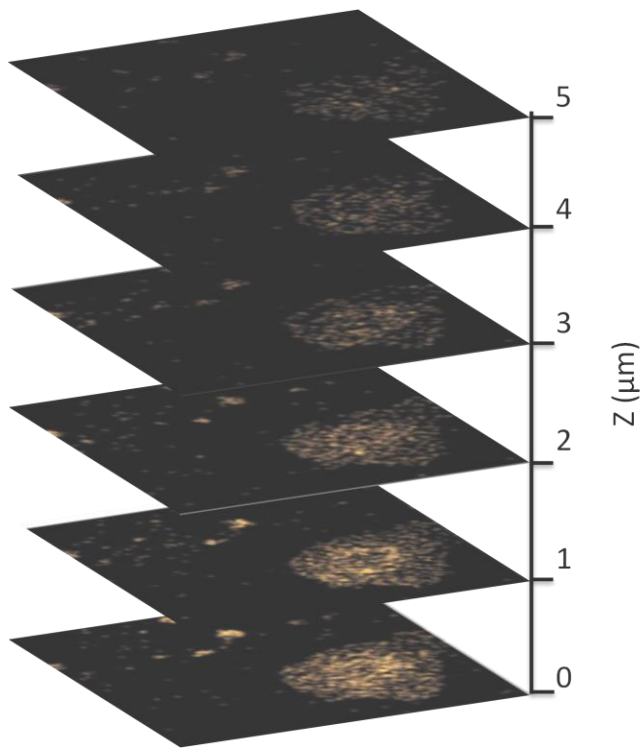

Figure 3: TPEF intensity raster scans performed at several Z positions of the droplets in heptane with TBA-Au NCs (size of the image : 637 x 637 µm). The overall time to perform each image was 1,24 s.